\documentclass[showpacs,twocolumn,amsmath,superscriptaddress]{revtex4-1}
\usepackage[dvipdfmx]{graphicx}
\usepackage{color,amsmath,lmodern,bm,array,braket,ulem}
\usepackage[italicdiff]{physics}

\definecolor{caribbeangreen}{rgb}{0.0, 0.8, 0.6}

\begin{document}

\title{Skyrmion-size dependence of the topological Hall effect: A real-space calculation
}%

\author{Akira Matsui}%
\email{matsui-akira897@g.ecc.u-tokyo.ac.jp}%
\affiliation{Department of Applied Physics, The University of Tokyo, Hongo, Bunkyo-ku, Tokyo, 113-8656, Japan}%
\author{Takuya Nomoto}%
\email{nomoto@ap.t.u-tokyo.ac.jp}%
\affiliation{Department of Applied Physics, The University of Tokyo, Hongo, Bunkyo-ku, Tokyo, 113-8656, Japan}%
\author{Ryotaro Arita}%
\affiliation{Department of Applied Physics, The University of Tokyo, Hongo, Bunkyo-ku, Tokyo, 113-8656, Japan}
\affiliation{RIKEN Center for Emergent Matter Science (CEMS), Wako 351-0198, Japan}
\date{\today}%

\begin{abstract}
Motivated by the recent discoveries of magnets harboring short-pitch skyrmion lattices,
we investigate the skyrmion-size dependence of the topological Hall effect.
By means of large-scale real-space calculations,
we find that the Hall conductivity takes its extreme value in the crossover region where both the real-space and momentum-space Berry curvature play a crucial role. 
We also investigate how the optimum skyrmion size ($\lambda_{\rm sk}^*$) , which separates the above two regions in the adiabatic region, depends on the lifetime of itinerant electrons ($\tau$) and coupling constant between electrons and localized spins ($J$). For the former, we show that $\lambda_{\rm sk}^{*}$ is proportional to $\sqrt{\tau}$, which indicates that $\lambda_{\rm sk}^{*}$ is much less sensitive to $\tau$ than the conventional expectation that $\lambda_{\rm sk}^{*}$ is proportional to the mean-free path $\propto \tau$. 
For the latter, we show that the non-adiabaticity considerably suppresses the topological Hall effect when the time scale determined by the skyrmion size and Fermi velocity is shorter than $1/J$. However, its effect on $\lambda_{\rm sk}^{*}$ is not so siginificant and $\lambda_{\rm sk}^{*}$ is about ten times the lattice constant in a wide range of $J$ and $\tau$.
\end{abstract}

\maketitle

\section{ Introduction}
A skyrmion is a vortex-like defect in helimagnets. Its non-trivial topology is characterized by the number of times spins wind around the unit sphere.
This skyrmion number governs numerous intriguing properties~\cite{NagaosaTokura,Back} that hold great promise for future spintronic applications~\cite{Fert2017,Zhang,Track}. Particularly, a small skyrmion is of great interest in that it is favorable for high-density data storage and high energy efficiency. Recently, short-pitch skyrmion lattices have became experimentally accessible~\cite{Kurumaji,Max,Khanh2020} and are attracting significant attention. Contrary to conventional skyrmion lattices in non-centrosymmetric materials induced by the Dzyaloshinskii–Moriya interactions~\cite{NagaosaTokura,Gayles2015,Koretsune2015,Kikuchi2016}, in these recently discovered centrosymmetric materials, small skyrmion lattices are stabilized by frustrations~\cite{Okubo,Leonov,BatistaLinHayami,nomoto} or four spin interactions originated from the higher-order interactions in itinerant magnets~\cite{Hayami,HayamiOzawaMotome,OzawaHayamiBarros,HayamiMotome}. 
 There, the skyrmion size is even comparable to the mean-free path of itinerant electrons~\cite{MaxNernst}. 
 
 Moreover, the recent study succeeded in tuning the skyrmion density systematically on the multilayers~\cite{Raju2021}. 
 The experimental accessibility to manipulate skyrmion lattices naturally motivates us to theoretically investigate how the skyrmion size affects the topological phenomena.


The non-trivial spin structure of skyrmions generates an emergent magnetic field acting on itinerant electrons. One notable manifestation of the emergent field is the THE, which was first experimentally discovered in MnSi~\cite{MnSi}. While  extensive theoretical efforts have been made so far for the THE~\cite{Nakazawa1,Nakazawa2,Denisov1,Denisov2,OnodaTataraNagaosa,TataraKawamura,NakazawaKohno,Rosales},
one of its interesting aspects is that we can obtain intuitive insights in the strong coupling limit. There, the effect of the non-collinear spin structure on itinerant electrons can be understood in terms of the Berry curvature $\Omega$ in real and momentum space. When the skyrmion size $\lambda_{\rm sk}$ is large, a real-space picture suggests that the topological Hall conductivity (THC) should behave as $\sim 1/\lambda_{\rm sk}^{2}$ since the strength of the real-space Berry curvature $\Omega(\vec{R})$ is proportional to the scalar spin chilarity or equivalently, the number of skyrmions in a given area~\cite{Bruno, Ye}.
Meanwhile, for small $\lambda_{\rm sk}$, analysis in momentum-space ($k$-space) is more useful: the THC is given by the sum of the $k$-space Berry curvature $\Omega(\vec{k})$ of occupied bands. Thus the THC becomes smaller when the skyrmion size becomes smaller since the size of the magnetic Brillouin zone (BZ) becomes larger and the number of occupied bands in the magnetic BZ becomes smaller~\cite{Hamamoto, Ohgushi, OnodaNagaosa, Batista}. 
 
This Letter focuses on the skyrmion-size dependence of the THE in the strong and medium coupling regimes. By changing $\lambda_{\rm sk}$, we can control the skyrmion density in the system, which determines the emergent magnetic field strengths. From the distinct behaviors of the THC in the limit of large and small $\lambda_{\rm sk}$, we expect that the THC takes its extreme values in the crossover region where both $\Omega(\vec{R})$ and $\Omega(\vec{k})$ play a crucial role. Throughout this letter, we focus on the optimum size of skyrmions ($\lambda_{\rm sk}^{*}$) for which the THC takes its extreme values because it is not merely a mathematical maximum but also indicates the crossover from the momentum to real-space description in the adiabatic region. To identify $\lambda_{\rm sk}^{*}$ for the THE, we systematically calculate the THC for a wide range of $\lambda_{\rm sk}$. We exploit the kernel polynomial method (KPM) to overcome the difficulty of growing computational costs for large skyrmion systems. The numerical calculation is particularly essential in the regions for which analytical approaches are not accessible. 

We first confirm that the THC depends on $\lambda_{\rm sk}$ non-monotonically as expected from the considerations for the limits of large and small $\lambda_{\rm sk}$. We then show that the THC takes its extreme value for different values of damping rates of itinerant electrons ($1/2\tau$). We naively expect that 
$\lambda_{\rm sk}^{*}$ is characterized by the mean-free path ($l$) which is proportional to $\tau$. However, we find that $\lambda_{\rm sk}^{*}$ is proportional to $\sqrt{\tau}$, which can be understood in terms of the band separations in the magnetic BZ. We further discuss the effect of non-adiabaticity by changing the coupling between electrons and spins ($J$), which becomes important when the time scale determined by the Fermi velocity ($v_F$) and $\lambda_{\rm sk}^{*}$ becomes shorter than $1/J$. We show that while the non-adiabaticity significantly affects the size of the THC, its effect on $\lambda_{\rm sk}^{*}$ is not so prominent. Therefore, in a wide range of $J$ and $\tau$, $\lambda_{\rm sk}^{*}$ is about ten times the lattice constant.

\section{Method }
The simplest model representing the interplay between itinerant electrons and localized spins is the double exchange model~\cite{AndersonHasegawa},
\[
H=\sum_{\langle i,j\rangle,\sigma} tc_{i,\sigma}^{\dagger}c_{j,\sigma} +J \sum_{i,\sigma,\sigma^{\prime}} (\vec{n}_{i}\cdot \vec{\sigma} )^{\sigma,\sigma^{\prime}}c_{i,\sigma}^{\dagger}c_{i,\sigma^{\prime}}, 
\]
where $J$ is the coupling between the electrons and spins. $\vec{n}_{i}$ stands for the normalized local spin at site $i$ which is treated as a purely classical quantity and the quantum fluctuation effects are neglected. We also neglect the spin-orbit coupling (SOC) in our model for simplicity. \footnote{
For large skyrmion systems, the SOC is small compared to the ferromagnetic interaction, and for small skyrmion systems such as Gd-based components, the SOC is negligible due to the orbital freezing. 
}
We calculate the THC for the triangular lattice and square lattice, where we assume triple-$Q$ and double-$Q$ skyrmion structure~\cite{NagaosaTokura}, 
\[
\vec{n}(\vec{r})\propto M\vec{e}_{z}+\sum_{i}\vec{m}_{i}\cos{(\vec{q}_{i}\cdot \vec{r}+\phi_{i})}+\vec{e}_{z}\sin{(\vec{q}_{i}\cdot\vec{r}+\phi_{i})},
\]
with $M=1$, $\phi_{i}=-\pi/2$,
$\vec{m}_{i}=\vec{e}_{z}\times \vec{q}_{i} /|\vec{e}_{z}\times \vec{q}_{i} |$, and $\vec{q}_{1}=2\pi/\lambda_{\rm sk} (1,0)$,
$\vec{q}_{2}=2\pi/\lambda_{\rm sk} (-1/2,\sqrt{3}/2)$, $\vec{q}_{3}=2\pi/\lambda_{\rm sk} (-1/2,-\sqrt{3}/2)$ for the triangular lattice, wheres $\vec{q}_{1}=2\pi/\lambda_{\rm sk} (1,0)$, $\vec{q}_{2}=2\pi/\lambda_{\rm sk} (0,1)$ for the square lattice. Hereafter, we set the hopping parameter $t=1$, the lattice constant $a=1$ and consider only the nearest-neighbor hoppings for simplicity. 

We calculate the THC ($\sigma_{xy}$) based on the linear response theory. We consider the effect of impurities by introducing the constant elastic scattering time $\tau$ and neglect the effect of vertex corrections. Although the vertex corrections play important roles in the weak coupling regimes~\cite{Nakazawa1,Nakazawa2}, they decrease as $\propto 1/(J\tau)$ in the strong to medium coupling regimes~\cite{Nakazawa1}. 
For non-interacting Hamiltonian, the conductivity can be obtained by the Kubo-Bastin formula~\cite{Bastin} and its Smrcka and Streda’s decomposition~\cite{Streda,Crepieux}. 
By substituting the non-interacting retarded and advanced Green functions with a constant finite lifetime $\tau$, i.e., $G^{\pm}(\varepsilon;x)=\int dx \delta (x-\hat{H})/(\varepsilon -x \pm i/2\tau)$ into the Kubo-Bastin formula, we obtain
\begin{equation}
\begin{split}
\sigma^{\alpha}_{\mu\nu} = \frac{e^{2}\hbar}{V}\int \dd x \dd y  F^{\alpha} (x,y) \mathrm{Tr} [ v_{\mu} \delta ( x - H ) v_{\nu}\delta ( y - H )  ], \label{sigma}
\end{split}
\end{equation} 
where $\alpha$ is either $\rm{I}$, $\rm{IIa}$, or $\rm{IIb}$, and 
\begin{equation}
\begin{split}
F^{\mathrm{I}} (x,y)  = - \frac{1}{2\pi }\frac{1}{x-y-i/\tau} \int \dd \epsilon \pdv{f}{\varepsilon}[g^{+}(\varepsilon;x)-g^{-}(\varepsilon;y)] \label{FI} \nonumber
\end{split}
\end{equation}
\begin{equation}
\begin{split}
F^{\mathrm{IIa}} (x,y)   = \frac{1}{2\pi} \frac{1}{x-y} \int \dd \epsilon  f(\varepsilon )  \mathrm{Im} [ g^{+}(\varepsilon;x)^{2}+g^{+}(\varepsilon;y)^{2} ] \label{FIIa} \nonumber
\end{split}
\end{equation}
\begin{equation}
\begin{split}
F^{\mathrm{IIb}} (x,y)   = \frac{1}{\pi}  \frac{1}{(x-y)^{2}}  \int \dd \epsilon f(\epsilon ) \mathrm{Im}[ g^{+}(\varepsilon;x)- g^{+}(\varepsilon;y) ],  \label{FIIb} \nonumber
\end{split}
\end{equation}
with $g^{\pm}(\varepsilon;x)=1/(\varepsilon-x\pm i/2\tau)$. Here, $f(\varepsilon)$ is the Fermi-Dirac distribution function and $v_{\mu}=iat/\hbar\sum_{i} ( c^{\dagger}_{i+\mu}c_{i}-c^{\dagger}_{i}c_{i+\mu} ) $ is the $\mu$-th component of the velocity operator. We divided the Fermi sea term into two terms, $\sigma_{xy}^{\rm{IIa}}$ and $\sigma_{xy}^{\rm{IIb}}$, following the convention in ~\cite{Kontani}. 
For large $\tau$, $\sigma_{xy}^{\rm{I}}$ and $\sigma_{xy}^{\rm{IIa}}$ cancel with each other, so that $\sigma_{xy} \sim \sigma_{xy}^{\rm{IIb}}$.
In this limit, the THC can be obtained by summing up $\Omega (\vec{k})$ for occupied bands (the Karplus-Luttinger formula)~\cite{KL,kSpace}. We can confirm this by taking the trace in Eq. (\ref{sigma}) explicitly for $\sigma_{xy}^{\rm IIb}$ using the eigenstates of the Hamiltonian and assuming that $\tau^{-1}$ in $g^{\pm}(\varepsilon;x)$ is infinitesimally small. Meanwhile, when $\tau^{-1}$ becomes larger than the typical band separation $\Delta$ in the magnetic BZ, $\sigma_{xy}^{\rm{IIb}}$ cancels with $\sigma_{xy}^{\rm{IIa}}$ and, therefore, the total conductivity reduces to $\sigma_{xy} \sim \sigma_{xy}^{\rm{I}}$.
In the actual calculation, we replace the derivative of the Fermi-Dirac distribution function with the Cauchy distribution, assuming that the temperature is sufficiently low (see Appendix A for details). 

To convert Eq.~\eqref{sigma} into a form convenient for KPM~\cite{KPM},
we use the following Chebyshev expansion of local density of states; $\mathrm{Tr} 
[\delta(x-\hat{H})]\cong\frac{1}{\pi\sqrt{1-x^{2}}}2\sum_{n=0}^{M_c}g_{n}h_{n}\mu_{n}T_{n}(x)$, where $M_{c}$ is the expansion order,  $h_{n}=1/(1+\delta_{n,0})$ is a normalization factor and $\mu_{n}=\mathrm{Tr}[T_{n}(\hat{H})]$ is the Chebyshev moment. Here, we introduce
the 
Jackson kernel $g_{n}$ to alleviate the Gibbs oscillations~\cite{Jackson1,Jackson2,KPM,WangBarros}. 
  
Substituting the delta function in the trace directly into Eq.~\eqref{sigma}, we obtain the following KPM formula for the conductivity,
\begin{equation}
\begin{split}
\sigma_{\mu\nu}^{\alpha}=\frac{e^{2}}{h}\frac{2\pi}{N} \sum_{m,n} h_{m}h_{n}g_{m}g_{n}C_{mn}^{\alpha}\mu_{mn,\mu\nu}, \label{KPM}
\end{split}
\end{equation} 
where $N$ is the number of sites, $C_{mn}^{\alpha}=\int \dd x \dd y  \frac{F^{\alpha} (x,y) }{\pi\sqrt{1-x^{2}}\pi\sqrt{1-y^{2}}}T_{m}(x)T_{n}(y)$, and $\mu_{mn,\mu\nu}=\mathrm{Tr}[\tilde{v}_{\mu}T_{m} (\hat{H})\tilde{v}_{\nu}T_{n} (\hat{H})]$ is the two-dimensional Chebyshev moment. Here, we introduced $\tilde{v}=\hbar v/a $. 
To reduce the numerical cost for the estimate of the trace, we exploit the random vector approximation~\cite{KPM}.
Note that $C_{mn}^{\alpha}$ can be efficiently calculated by the Chebyshev Gauss quadrature~\cite{KPM,WangBarros,BarrosKato}.

\section{Results and Discussions} 
Using the model and method described above, we compute the conductivity for various $\lambda_{\rm sk}$. Throughout this paper, we set the system size $N=192^{2}$ and the number of random vectors $R=100$. The tunable parameters are coupling constant $J$, damping rate $\tau^{-1}$, and chemical potential $\mu$. $M_c=5000$ is enough to achieve the convergence of $\sigma_{xy}$ for all $\tau^{-1}$ when the skyrmion size $\lambda_{\rm sk}$ is smaller than 96.

\begin{figure}[t]
    \centering
    \includegraphics[width=0.45\textwidth]{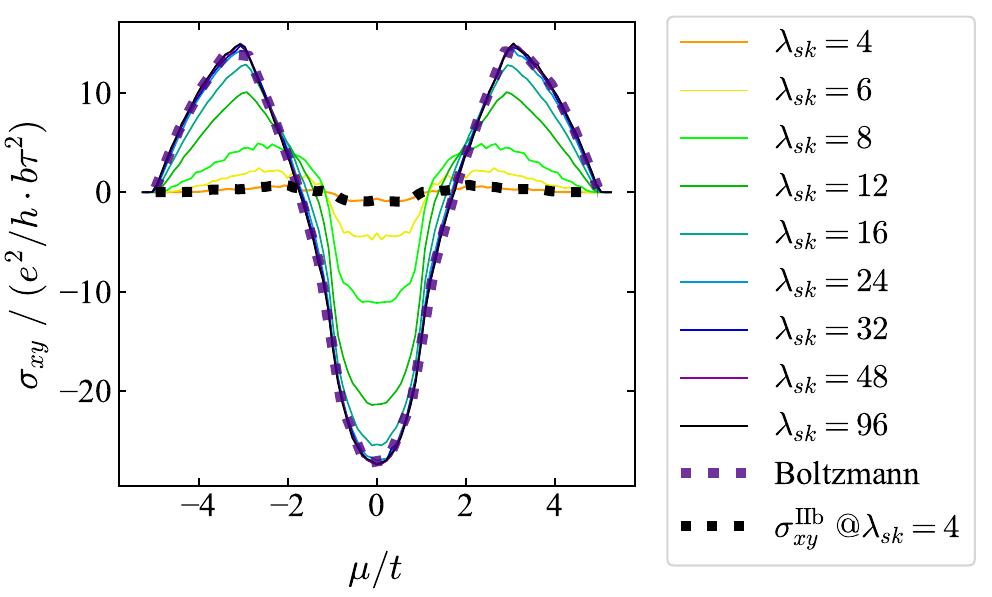}
     \caption{ 
     $\mu$-dependence of the topological Hall conductivity ($\sigma_{xy}$) for $\tau^{-1}=0.1$ in the unit of $e^{2}/h\cdot b^{2}\tau^{2}$. For $\lambda_{\rm sk}>16$, the results show good agreement with that for a free-electron system under a uniform magnetic field $b$. For $\lambda_{\rm sk}=4$, $\sigma_{xy}^{\rm IIb}$ calculated from the $k$-space Berry curvature $\Omega({\vec k})$ for occupied bands agrees well with $\sigma_{xy}$.}
     \label{fig:1}
\end{figure}

Let us first discuss the $\mu$-dependence of the conductivity for the square 
lattice. In Fig.~\ref{fig:1}, we show the THC for various $\lambda_{\rm sk}$ with $J=1.0$ and $\tau^{-1} = 0.1$. We compare the results with that for a free-electron system under a uniform magnetic field $b=(e/h)/\lambda_{\rm sk}^{2}$ calculated with the linearized Boltzmann transport theory:
\begin{align}
\sigma_{xy}^{\rm B}&=\frac{e^2}{h}b\tau^{2}
\sum_{\sigma}\sigma \int \frac{d^{d}k}{(2\pi)^{d}}(-\frac{df(\varepsilon_{\sigma})}{d\varepsilon})
\nonumber \\ &\hspace{2cm}\times  (v_{x\sigma}^{2}m_{yy\sigma}^{-1}-v_{x\sigma}v_{y\sigma}m_{xy\sigma}^{-1}), 
\label{boltz}
\end{align}
where $\sigma=1~(-1)$ for the majority (minority) spin, and $v_{i\sigma}=\partial \varepsilon_{\sigma}/\partial k_{i}$, $m_{ij\sigma}^{-1}=(1/\hbar^{2})\partial^{2} \varepsilon_{\sigma} /\partial k_{i} \partial k_{j}$. 
This comparison indicates that the system can be mapped to a free-electron system when $\lambda_{\rm sk}$ is not smaller than $24$.
However, this real-space description is not valid for $\lambda_{\rm sk}\le 16$. In Fig.~\ref{fig:1}, we compare $\sigma_{xy}$ and $\sigma_{xy}^{\rm{IIb}}$ for $\lambda_{\rm sk}=4$. We see that the agreement is remarkably well, indicating that the main contribution to the THE can be described in terms of $\Omega({\vec k})$ rather than $\Omega({\vec R})$.

\begin{figure}[t!]
    \centering
    \includegraphics[width=0.45\textwidth]{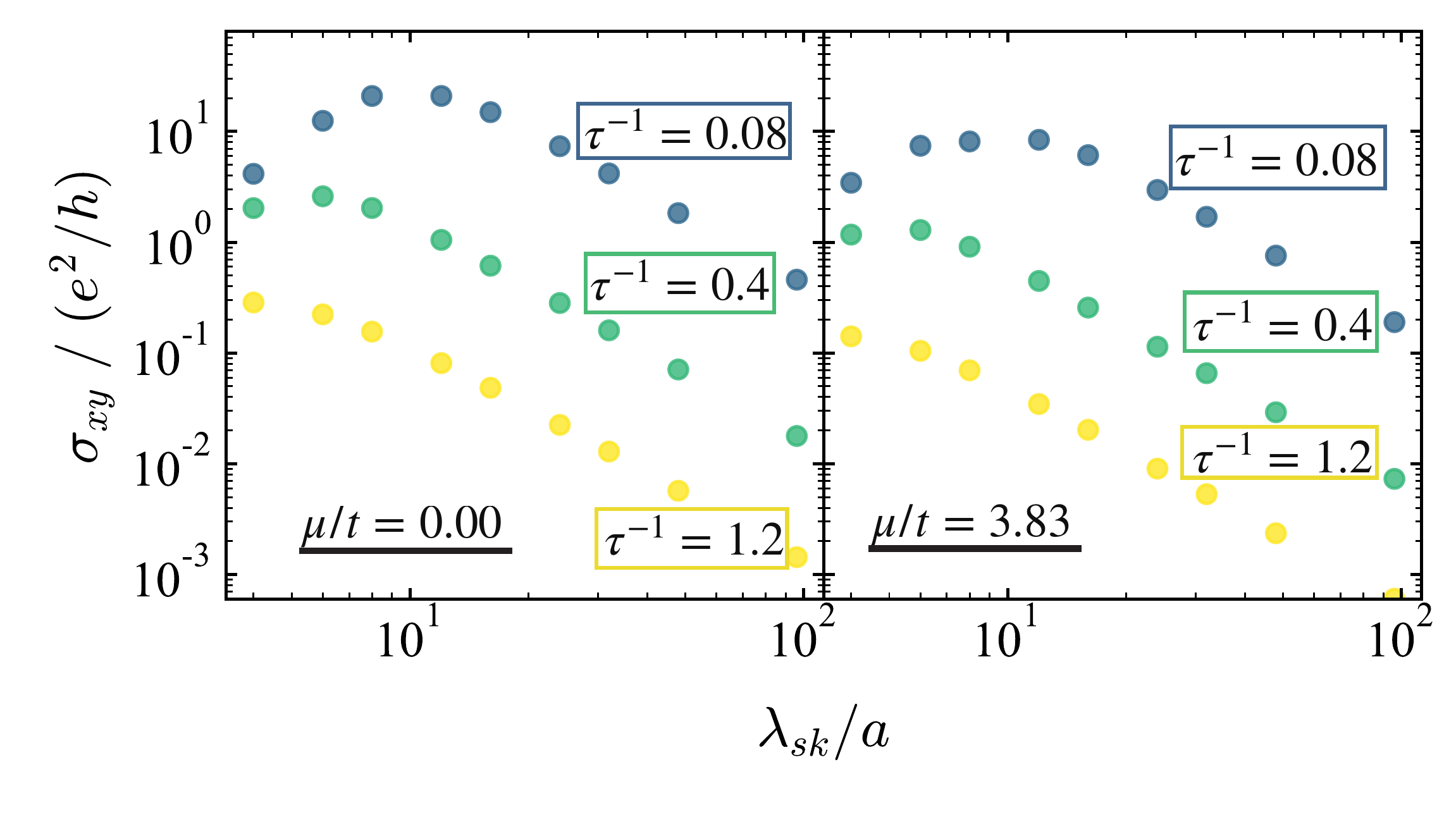}
     \caption{ 
     $\lambda_{\rm sk}$-dependence of the conductivity for $\tau^{-1}=0.08,~0.4,~1.2$. The chemical potential $\mu$ is $0.00$ and $3.83$ for the left and right panels, respectively. 
     }
    \label{fig:2}
\end{figure}

To identify the optimum skyrmion size ($\lambda_{\rm sk}^*$) for the THC, we investigate the $\lambda_{\rm sk}$-dependence of $\sigma_{xy}$. As shown in Fig.~\ref{fig:2}, $\sigma_{xy}$ is not a monotonic function of $\lambda_{\rm sk}$ and has a peak at $\lambda_{\rm sk}=\lambda_{\rm sk}^*$. While $\lambda_{\rm sk}^*$ obviously depends on $\tau^{-1}$, its $\mu$-dependence is not so significant. According to the real-space description~\cite{Bruno, Ye}, the conductivity should be promotional to $1/\lambda_{\rm sk}^{2}$ as the emergent field is stronger for denser skyrmion lattices~\cite{Nakazawa1}. 

Note that the non-monotonic behavior with respect to $\lambda_{\rm sk}$ obtained from our calculation is distinct from the non-monotonic temperature dependence~\cite{Fujishiro2021,Raju2021} as we are not considering the effect of the chiral spin fluctuations.

For smaller $\lambda_{\rm sk}$, on the contrary, the conductivity increases as the skyrmion size becomes large. This is consistent with the $k$-space description since $\sigma_{xy}$ increases with the growing number of bands in the magnetic BZ~\cite{Hamamoto,Gobel1}. Such behaviors are observed recently for $J/t \ll 1$ and small $\lambda_{\rm sk}$~\cite{Batista}. 

\begin{figure}[t]
    \centering
    \includegraphics[width=0.45\textwidth]{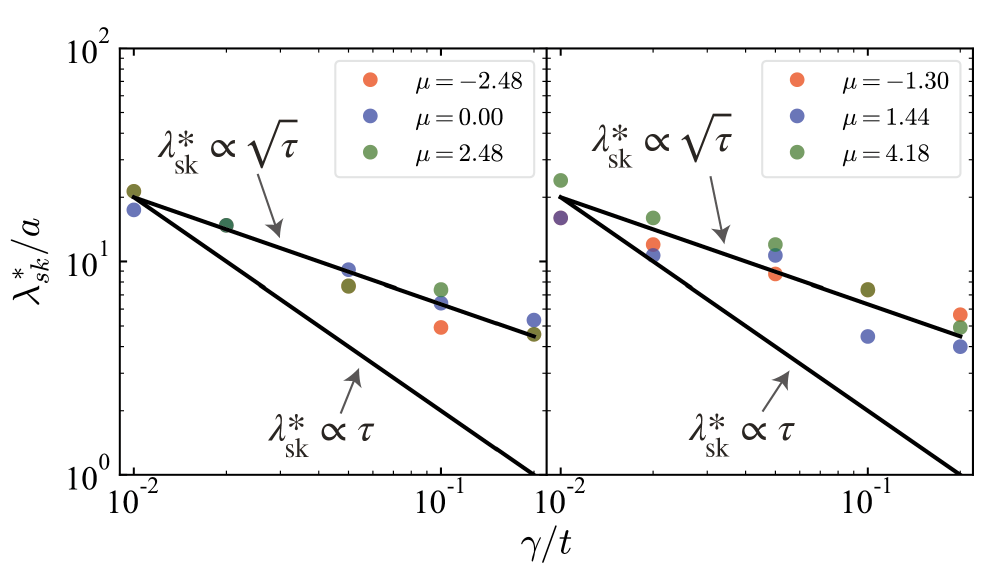}
    \caption{Optimum $\lambda_{\rm sk}^{*}$ as a function of $\tau^{-1}$ for various $\mu$'s. The left and right panels correspond to the square and triangular lattices, respectively. The solid lines indicate $\lambda_{\rm sk} \propto \sqrt{\tau}$ and $\lambda_{\rm sk} \propto \tau$.}
     \label{fig:4}
\end{figure}

\begin{figure}[t]
    \centering
    \includegraphics[width=0.45\textwidth]{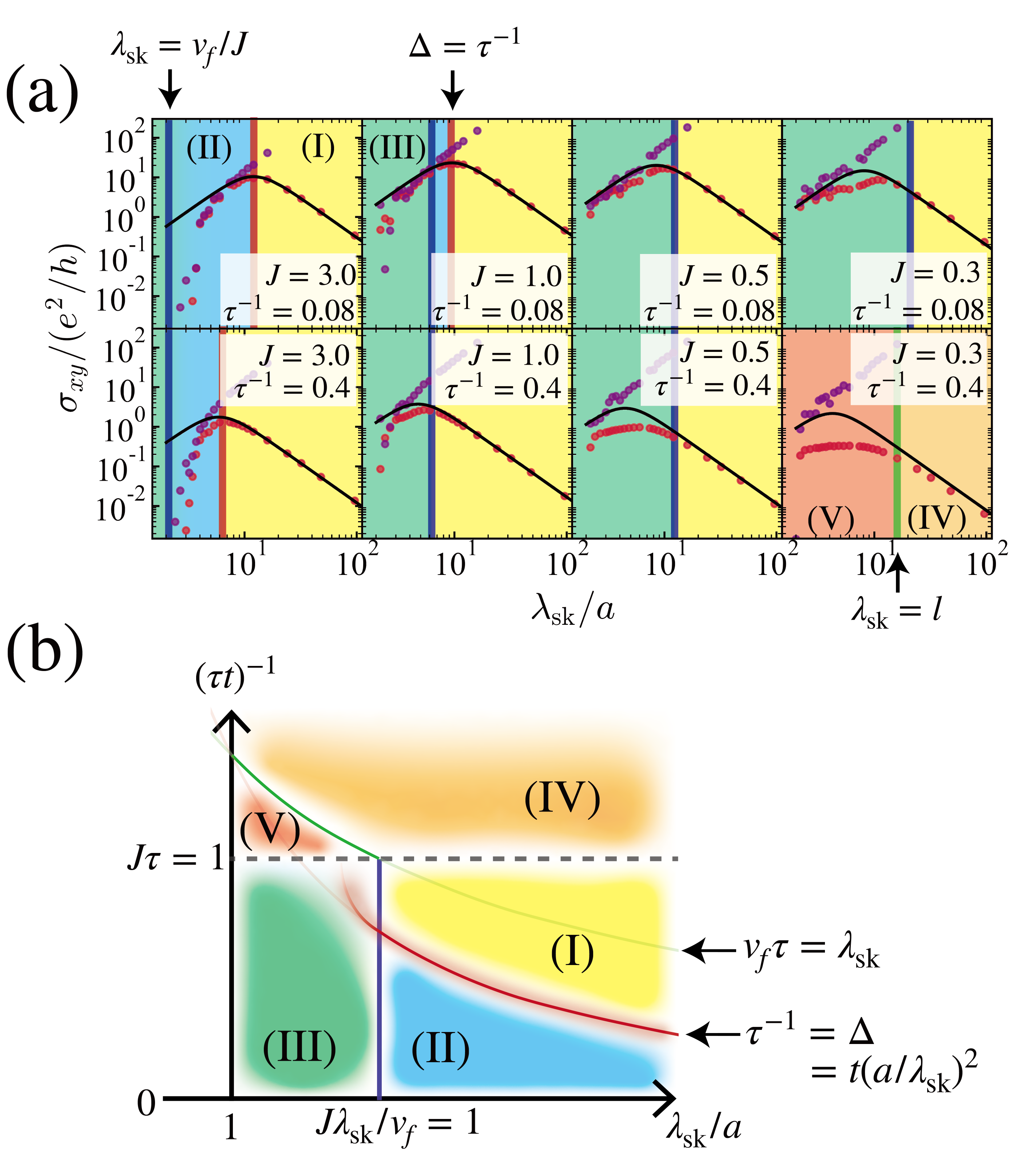}
     \caption{(a) The $\lambda_{\rm sk}$-dependence of the Hall conductivity $\sigma_{xy}$ (red circles) and $\sigma_{xy}^{\rm IIb}$ (purple circles) for $\tau^{-1}=0.08$ (the upper four) and for $\tau^{-1}=0.4$ (the lower four). From left to right, $J$ is $3.0,~1.0,~0.5,$ and $0.3$, respectively. The blue, red and green solid lines indicate $\lambda_{\rm sk}= v_f/J$, $\Delta=\tau^{-1}$, and $\lambda_{\rm sk}=l$, respectively. (b) Five types of $\lambda_{\rm sk}$-dependence of $\sigma_{xy}$ in the plane of $\tau^{-1}$ and $\lambda_{\rm sk}$. The blue solid line ($\lambda_{\rm sk}= v_f/J$) and gray dotted line ($J=\tau^{-1}$) determine the boundary between the adiabatic and non-adiabatic regime.}
     \label{fig:3}
\end{figure}

In Fig.~\ref{fig:4}, we plot $\lambda_{\rm sk}^*$ as a function of $\tau^{-1}$ for the square lattice and triangular lattice. When the mean-free path $l$ is smaller than $\lambda_{\rm sk}$, the real-space description is valid~\cite{Nakazawa1}, which suggests $\lambda_{\rm sk}^*\propto \tau$. However, we find $\lambda_{\rm sk}^* \propto \sqrt{\tau}$ for all $\mu$, which can be explained by comparing $\tau^{-1}$ and averaged band separation $\Delta$ in the magnetic BZ 
as follows. $\Delta$ can be estimated in analogy to the formation of the Landau level~\cite{OnodaTataraNagaosa}, namely, $\Delta  \sim N/\lambda_{\rm sk}^{2} /  \mathrm{DOS} \sim t(a/\lambda_{\rm sk})^{2}$ with $N$ denoting the number of sites in the system. 

As we have mentioned in the method section, the $k$-space description for $\sigma_{xy}$ is valid, i.e., $\sigma_{xy}\sim\sigma_{xy}^{\rm IIb}$ only when the typical energy difference in the magnetic BZ $\Delta$ is larger than $\tau^{-1}$. Indeed, the monotonic increase of the THC observed in this region is consistent with the result of the calculations without the effect of impurities~\cite{Gobel1,Hamamoto}. On the other hand, for $\Delta <\tau^{-1}$, $\sigma_{xy}$ reduces to $\sigma_{xy}^{\rm I}$ and does not necessarily monotonically increase as a function of $\lambda_{\rm sk}$. 
Note that, in the large $\lambda_{\rm sk}$ limit, the monotonic decrease of the THC observed in our calculation is consistent with the monotonic decrease of the resistivity $\rho_{yx}\cong \sigma_{xy}/\sigma_{xx}^{2}$.
Since $\Delta\propto\lambda_{\rm sk}^{-2}$, we get the simple relation $\lambda_{\rm sk}^{*}\propto \sqrt{\tau}$. 
We further note that we can also estimate $\lambda_{\rm sk}^*$ from  $(e\lambda_{\rm sk}^{2}/h)\sim \mu_{\rm{mob}}$, where $\mu_{\mathrm{mob}}$ is the mobility of itinerant electrons. This relation will be an useful indicator to see whether a real-space or $k$-space description is more appropriate to understand the experimental data. 

Up to this point, the coupling constant $J$ has been fixed to be $1.0$, for which the itinerant electrons adiabatically interact with spins. However, for small $J$, we can no longer map the system to a free-electron system under a uniform magnetic field even for large $\lambda_{\rm sk}$~\cite{Metalidis, Stern}. Besides, if the energy scale of $J$ is much smaller than $\tau^{-1}$, the Karplus-Luttinger formula also becomes inaccurate even when $\Delta$ is sufficiently smaller than $\tau^{-1}$. Thus, when the electron-spin coupling is not adiabatic, both the real-space and $k$-space description for $\sigma_{xy}$ discussed above lose its validity. 
Whether the electron-spin coupling is adiabatic or not is determined by the time scale of $1/J$:
When the mean-free path ($l=v_f\tau$, where $v_f$ is the Fermi velocity) is smaller than $\lambda_{\rm sk}$, $1/J$ must be shorter than $\tau$. Otherwise, $1/J$ must be shorter than $\lambda_{\rm sk}/v_f$~\cite{Metalidis}. The adiabatic criterion $J\lambda_{\rm sk}/v_{f}$ is also implied from the scattering theory analysis as the condition for the validity of the Born approximation.~\cite{Denisov3}

In Fig.~\ref{fig:3}(a), keeping the condition $\Delta > \tau^{-1}$, we show how the $\sigma_{xy}$-$\lambda_{\rm sk}$ curve changes in the non-adiabatic regime. 
For the upper and lower four panels, $\tau^{-1}$ is set to be 0.08 and 0.4, respectively. 
From left to right, $J$ is 3.0, 1.0, 0.5, and 0.2. Except for the case of $J$=0.2 and $\tau^{-1}$=0.4 (the right bottom panel), the condition $J>\tau^{-1}$ is satisfied. Thus, whether the system is adiabatic or not is determined by the condition $\lambda_{\rm sk}=v_f/J$ indicated by the blue solid lines. In the right bottom panel, the green line denotes $\lambda_{\rm sk}=l$, i.e., whether the mean-free path is larger than the skyrmion size.  

We then compare the results in Fig.~\ref{fig:3}(a) with a simple formula derived from the Drude theory for a free-electron system under an effective magnetic field $b\propto 1/\lambda_{\rm sk}^{2}$:  $\sigma = K/(\lambda_{\rm sk}^{2}+\alpha^{2})$. 
Here, $K$ is determined so that $\sigma$ agrees with $\sigma_{xy}^{\rm B}$ calculated from the linearized Boltzmann transport theory for $\lambda_{\rm sk} \gg \alpha$, and $\alpha$ is tuned so that $\sigma$ takes its extreme value at the point where $\sigma^{\rm B}_{xy}$ and $\sigma^{\rm IIb}_{xy}$ intersect.
The result is shown with the solid black line in each panel. 
When $\lambda_{\rm sk}$ is larger than $v_f/J$ or $l$, the $\lambda_{\rm sk}$-dependence of $\sigma_{xy}$ is successfully reproduced by the simple model.
However, there are significant deviations for small $\lambda_{\rm k}$ even in the adiabatic regime when the chemical potential lies near the edge of the band (see the upper left panel of Fig.~\ref{fig:3}(a)).



Except for the right bottom panel, we see that the non-adiabatic regime expands as $J$ decreases. It should be noted that even in the non-adiabatic regime, the Karplus-Luttinger formula ($\sigma_{xy}^{\rm IIb}$, purple circles) gives reliable results for small $\lambda_{\rm sk}$, as far as the condition $\Delta>\tau^{-1}$ is satisfied. Interestingly, while the maximum value of $\sigma_{xy}$ (red circles) is suppressed in the non-adiabatic regime, $\lambda_{\rm sk}^{*}$ does not change drastically. Namely, for both $\tau^{-1}=$0.08 and 0.4, $\lambda_{\rm sk}^*$ is always $\sim$ 10. We see the growing non-adiabatic contribution even in the spin-resolved Hall conductivity calculation (see Appendix B for details). According to the scattering theory analysis, the consequence of the non-adiabaticity is the crossover from the spin Hall effect to the charge Hall effect.~\cite{Denisov3} Indeed, we observe the same tendency, which results in the suppression discussed above.

In Fig.~\ref{fig:3}(b), we summarize the different types of $\lambda_{\rm sk}$-dependence in the plane of $\lambda_{\rm sk}$ and $\tau^{-1}$.
The below is a detailed description of each region.


    {\bf Region I}: The real-space picture holds here, namely, $\sigma_{xy}\propto \int \dd^{2} R \Omega(\vec{R}) \propto 1/\lambda_{\rm sk}^2$. 
    The Hall conductivity takes its extreme value when $\Delta=\tau^{-1}$ which separates Region I and II (the red shaded line). 
    
    {\bf Region II}: The momentum-space picture holds. The Hall conductivity can be calculated by the Karplus-Luttinger formula; $\sigma_{xy} \propto \int \dd^{2} k \Omega(\vec{k})\propto \lambda_{\rm sk}^2$.
    
    {\bf Region III}: The non-adiabaticity is non-negligible in this region. Although the Hall conductivity is suppressed compared to the adiabatic regions (Region I and II) due to the non-adiabaticity, the Hall conductivity can be well estimated by the Karplus-Luttinger formula in the clean limit, i.e., $\sigma_{xy} \propto \int \dd^{2} k \Omega(\vec{k})\propto \lambda_{\rm sk}^2$. The optimum skyrmion size is also affected, yet the deviation is not significantly large. 
    
    {\bf Region IV and V}: This region corresponds to the small-$J$ limit ($J\tau<1$) where the non-adiabaticity is dominant. Due to the non-adiabaticity, the real-space picture fails even for the large skyrmion size limit. In Region V, the $\lambda_{\rm sk}$-dependence of $\sigma_{xy}$ is considerably suppressed, which is consistent with the previous perturbative study.~\cite{Nakazawa2} Note that the effect of the vertex corrections is not negligible in this region unlike the others.~\cite{Nakazawa1}

In real materials, the $\rm Gd_{2}PdSi_{3}$ is reported to have $\lambda_{\rm sk}^{*}/\lambda_{\rm sk}\sim \sqrt{\mu_{\rm mob}b} \gtrsim 0.28$~\cite{MaxNernst}, while the other Gd-based skyrmion host materials such as $\rm Gd_{3}Ru_{4}Al_{12}$~\cite{Max} and $\rm GdRu_{2}Si_{2}$~\cite{Khanh2020} are cleaner compared to $\rm Gd_{2}PdSi_{3}$, suggesting the possible crossover occurring. 


\section{Conclusion} 
In the present study, we have investigated the $\lambda_{\rm sk}$-dependence of the THE. While extensive analytical calculations have been performed so far for different parameter regions, we have performed a large-scale real-space numerical calculation. We determined the optimum $\lambda_{\rm sk}^*$ for the THE and discussed how it depends on parameters such as the elecqtron-spin coupling $J$ and electron lifetime $\tau$. For the $\tau$-dependence of $\lambda_{\rm sk}^*$, we found that $\lambda_{\rm sk}^{*}$ is proportional to $\sqrt{\tau}$ but not to $\tau$ (i.e., mean-free path). 
This behavior can be understood in terms of the band separation in the magnetic BZ.
For the $J$-dependence, we found that while the non-adiabticity suppresses the THE considerablly, its effect on $\lambda^*_{\rm sk}$ is not so significant. Therefore, the size of $\lambda^*_{\rm sk}$ is about ten times the lattice constant in a wide range of $J$ and $\tau$.

\section*{ Acknowledgements}
We would like to thank K. Nakazawa for fruitful discussion. We acknowledge the Center for Computational Materials Science, Institute for Materials Research, Tohoku University for the use of MASAMUNE-IMR (MAterials science Supercomputing system for Advanced MUltiscale simulations towards NExt generation). (Project No. 202012-SCKXX-0009) This work was supported by a Grant-in-Aid for Scientific  Research (No. 19K14654, No. 19H05825, No. 20K21067, No. 21H04437, and No. 21H04990), “Program for Promoting Researches on the Supercomputer Fugaku” (Project ID: hp200132) from MEXT, JST-CREST (JPMJCR18T3), JST-PRESTO (JPMJPR20L7) and JST-Mirai Program (JPMJMI20A1).

\appendix
\section{The comparison of the Cauchy distribution and the Fermi-Dirac distribution}
\begin{figure}[hb]
    \centering
    \includegraphics[width=0.45\textwidth]{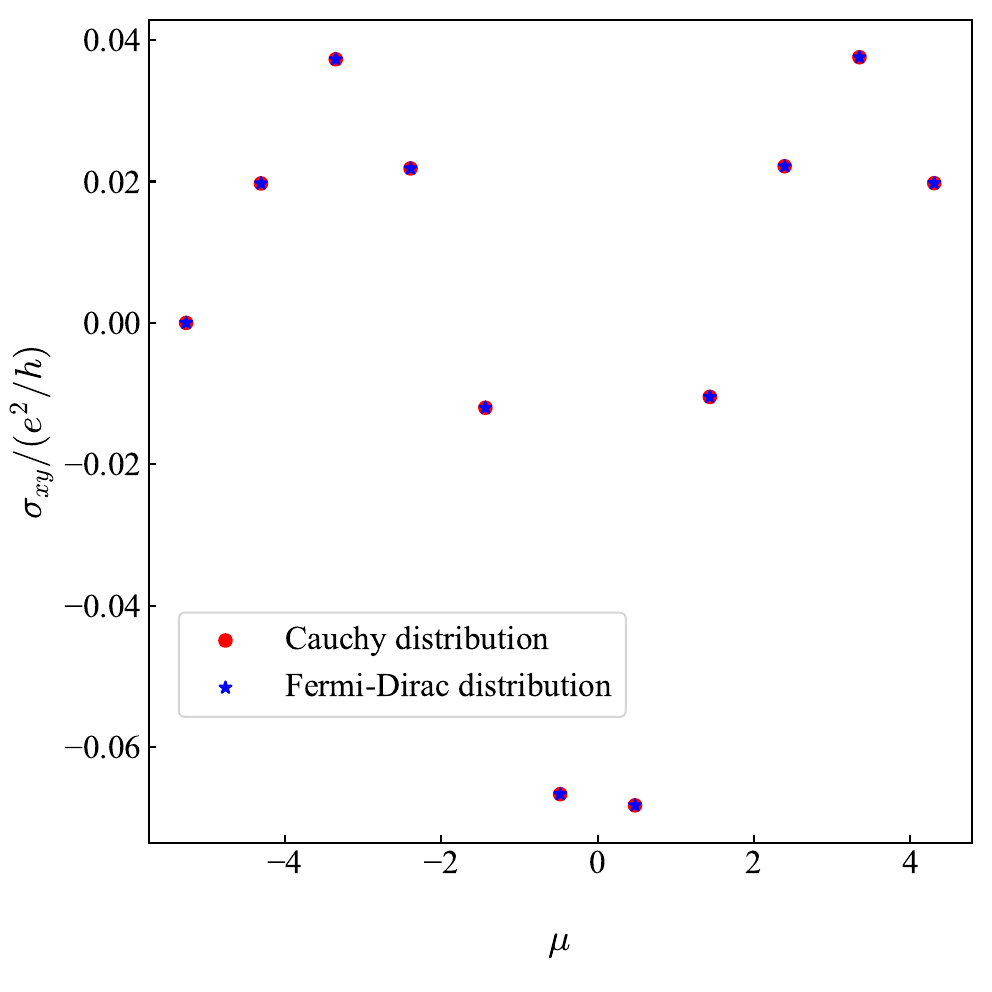}
     \caption{ The comparison of the results of the calculation using the two distributions: the Cauchy distribution and the Fermi-Dirac distribution.}
     \label{fig:ap1}
\end{figure}

In the calculation of the Hall conductivity, we used the Cauchy distribution instead of the Fermi-Dirac distribution to reduce the computational cost. Here, we present the validity of the replacement by calculating the Hall conductivity using the above two distributions. Figure~\ref{fig:ap1} shows the results of the calculations, where the parameters are set as follows: $L=192,~\lambda_{\rm sk}=96,~J=1.0,~\tau^{-1}=0.2,~M=2000,~r=20$. It shows the good agreement of the two methods and the difference is negligible in the scale of the figure.

\section{The crossover from the spin Hall effect to the charge Hall effect}
\begin{figure}[t]
    \centering
   \includegraphics[width=0.45\textwidth]{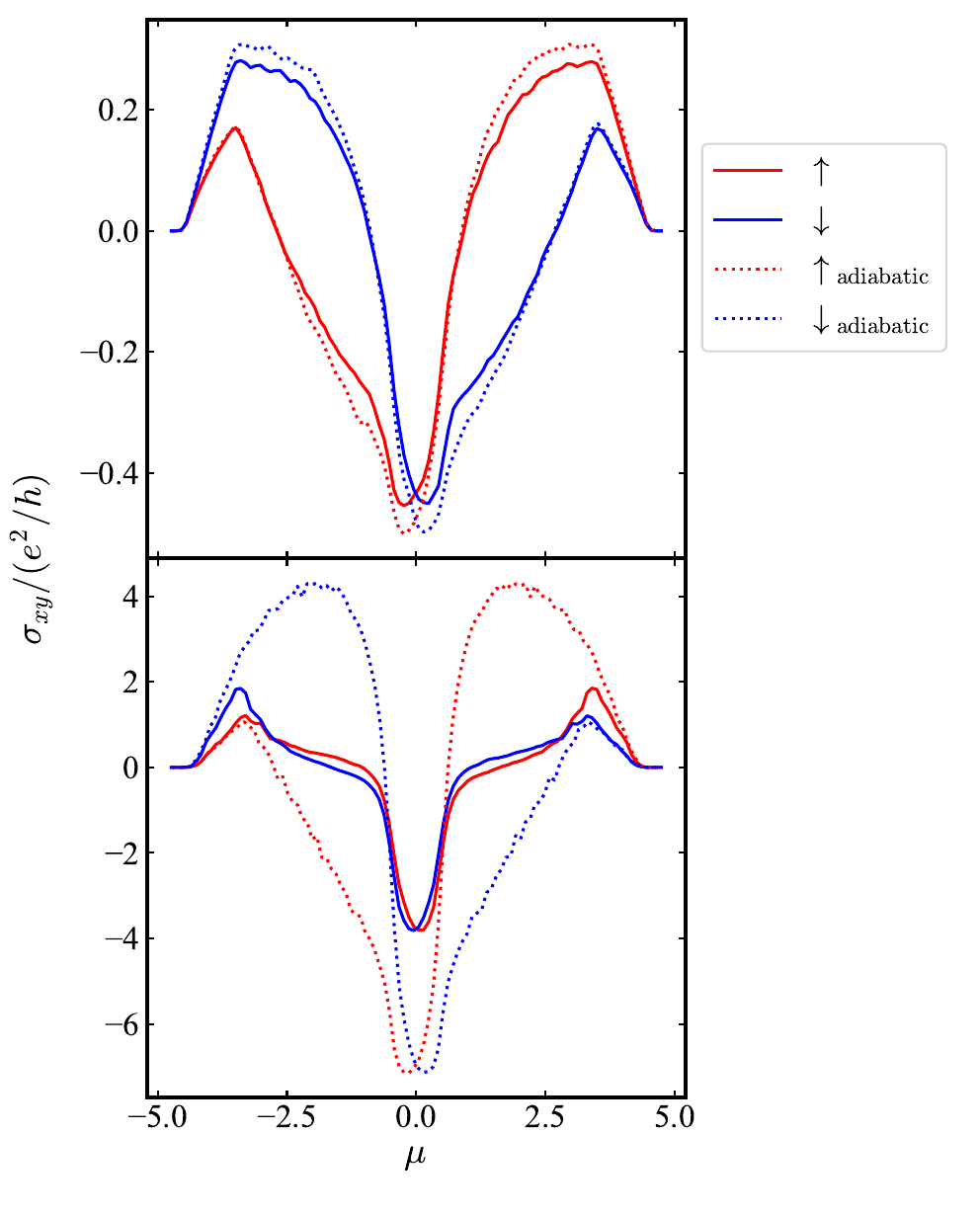}
     \caption{  Spin-resolved Hall conductivity calculated for the full Hamiltonian (solid line) and the adiabatic Hamiltonian (dotted line) for $\lambda_{\rm sk}=32$ (upper panel) and $\lambda_{\rm sk}=8$ (lower panel).}
     \label{fig:ap2}
\end{figure}

As mentioned in the main text, the charge Hall to spin Hall transition presented previously~\cite{Denisov3} is reproduced in our calculation. In order to see the transition, we present here the comparison of the up spin current conductivity and the down spin current conductivity in Fig.~\ref{fig:ap2}. The calculation is carried out for $J=0.5$, $\tau^{-1}=0.2$, and we define the up and down spin current operators by the spin indices measured in the global frame, not the local frame as below.
\begin{align}
\vec{J}_{\sigma}= \frac{\vec{J}P_{\sigma}+P_{\sigma}\vec{J}}{2} ,
\end{align}
where $P_{\sigma}$ is the projection operator onto the subspace of up or down spin in the global frame. Figure~\ref{fig:ap2} shows the up and down spin channels exhibit the opposite signs to each other for $\lambda_{\rm sk}=32$, while they have the same signs in most of the energy region for $\lambda_{\rm sk}=8$.  We further calculate the conductivity using the adiabatic Hamiltonian and compare the result. The adiabatic component of an operator $A$ is obtained as below.
\begin{align}
A_{\rm ad}=\frac{A+QAQ}{2}, 
\end{align}
where $\Braket{i,\sigma_{\rm local}|Q|j,\sigma^{\prime}_{\rm local}}=(-1)^{\sigma_{\rm local}}\delta_{i,j}\delta_{\sigma_{\rm local},\sigma_{\rm local}^{\prime}}$. Here $\sigma_{\rm local}$ denotes the spin indices measured in the local frame. The result is shown in Fig.~\ref{fig:ap2} by the dotted line. The comparison between the total spin-resolved conductivity and its adiabatic contribution reveals that the non-adiabaticity becomes dominant for $\lambda_{\rm sk}=8$, which is in good agreement with the growing contribution of the off-diagonal components of the scattering matrix in the scattering theory analysis.~\cite{Denisov3} As discussed in the main text, we can see the non-negligible effect of the non-adiabaticity for smaller values of $\lambda_{\rm sk}$.

\bibliography{bib2}

\end{document}